\documentclass[10pt, conference]{IEEEtran}
\IEEEoverridecommandlockouts
\usepackage{cite}
\usepackage{amsmath,amssymb,amsfonts}
\usepackage{dsfont}
\usepackage{algorithmic}
\usepackage{graphicx}
\usepackage{textcomp}
\usepackage{xcolor}
\def\BibTeX{{\rm B\kern-.05em{\sc i\kern-.025em b}\kern-.08em
    T\kern-.1667em\lower.7ex\hbox{E}\kern-.125emX}}

\usepackage{mathtools}
\DeclarePairedDelimiter\bra{\langle}{\rvert}
\DeclarePairedDelimiter\ket{\lvert}{\rangle}
\usepackage{subcaption}
\usepackage{hyperref}

\begin{document}

\title{Gaussian Boson Sampling for binary optimization
    \thanks{
        DOI: \href{https://doi.ieeecomputersociety.org/10.1109/QCE57702.2023.10268}{10.1109/QCE57702.2023.10268}

        © 2023 IEEE. Personal use of this material is permitted. Permission from IEEE must be obtained for all other uses, in any current or future media, including reprinting/republishing this material for advertising or promotional purposes, creating new collective works, for resale or redistribution to servers or lists, or reuse of any copyrighted component of this work in other works.

        \textbf{Acknowledgements.} This work is funded by the German Federal Ministry of Education and Research (BMBF) under the project "PhoQuant". This work is supported by the European Union’s Horizon Europe Framework Programme (HORIZON) under the ERA Chair scheme with grant agreement no.\ 101087126. This work is supported with funds from the Ministry of Science, Research and Culture of the State of Brandenburg within the Centre for Quantum Technologies and Applications (CQTA). J.C. and T.S. are grateful for valuable feedback from the Simulation and Software team at Q.ANT GmbH.}
}

\author{\IEEEauthorblockN{1\textsuperscript{st} Jean Cazalis}
    \IEEEauthorblockA{\textit{Q.ant GmbH} \\
        Stuttgart, Germany \\
        jean.cazalis@qant.gmbh}
    \and
    \IEEEauthorblockN{2\textsuperscript{nd} Yahui Chai}
    \IEEEauthorblockA{\textit{CQTA} \\
        \textit{DESY}\\
        Zeuthen, Germany \\
        yahui.chai@desy.de}
    \and
    \IEEEauthorblockN{3\textsuperscript{rd} Karl Jansen}
    \IEEEauthorblockA{\textit{CQTA} \\
        \textit{DESY}\\
        Zeuthen, Germany \\
        karl.jansen@desy.de}
    \and
    \IEEEauthorblockN{4\textsuperscript{th} Stefan Kühn}
    \IEEEauthorblockA{\textit{CQTA} \\
        \textit{DESY}\\
        Zeuthen, Germany \\
        stefan.kuehn@desy.de}
    \and
    \IEEEauthorblockN{5\textsuperscript{th} Tirth Shah}
    \IEEEauthorblockA{\textit{Q.ant GmbH} \\
        Stuttgart, Germany \\
        tirth.shah@qant.gmbh}
}

\maketitle

\begin{abstract}
    In this study, we consider a Gaussian Boson Sampler for solving a Flight Gate Assignment problem. We employ a Variational Quantum Eigensolver approach using the Conditional Value-at-risk cost function. We provide proof of principle by carrying out numerical simulations on randomly generated instances.
\end{abstract}


\section{Introduction}

The Variational Quantum Eigensolver (VQE), a variational approach for finding the ground state energy of Hamiltonians, makes use of the quantum device to prepare an ansatz state in the form of a parametric quantum circuit. This ansatz can be represented by $\boldsymbol{\theta}\in\mathbb{R}^p \mapsto \ket{\varphi(\boldsymbol{\theta})}$. Then, the goal is to minimize the Hamiltonian expectation value in a hybrid quantum-classical loop. VQE has recently been proposed to address combinatorial optimization problems~\cite{Nannicini2019}.

In this paper, we explore a VQE-based approach, already presented in~\cite{Chai2023}, to address the Flight-Gate Assignment (FGA) problem. As in~\cite{Banchi2020}, we use a Gaussian Boson Sampler (GBS) instead of a digital quantum computer. In contrast, our approach employs a more general GBS ansatz, along with a cost function that
is computed analytically in the case of VQE.

\section{Gaussian boson sampling}

A GBS is a non-universal photonic quantum platform and one of the prominent candidates for demonstrating quantum advantage in the near future~\cite{Madsen2022, Zhong2020}.

The GBS ansatz consists of pure $N$-modes Gaussian states without displacement sampled using threshold detectors (TD) \cite{Quesada2018}. The quantum state is prepared from the vacuum by applying squeezing gates with parameters $(r_1,\dots,r_N) \in \mathbb{R}_+^N$ followed by a universal passive interferometer, whose action on creation/annihilation operators is described by a $N\times N$ unitary $U$. The output state is characterized by its Husimi covariance matrix $\Sigma$. Then, the photons are sent to the TDs which record a 'click' when one or more photons are detected, 'no click' otherwise. These outcome measures are described by the measurement operators
\begin{align*}
    \hat{\Pi}^{(0)}_j = \ket{0_j}\bra{0_j} \quad \text{and} \quad \hat{\Pi}^{(1)}_j = \mathbb{I} - \ket{0_j}\bra{0_j}\, ,
\end{align*}
where $\mathbb{I}$ is the identity operator and $\ket{0_j}$ the vacuum on the optical mode $j$. The probability to detect a pattern is proportional to the \emph{Torontonian} of the submatrix of $O = \mathbb{I} - \Sigma^{-1}$ corresponding to the pattern.

\begin{figure*}[ht]
    \centering
    \begin{minipage}{0.45\linewidth}
        \centering
        \includegraphics[width=0.97\linewidth]{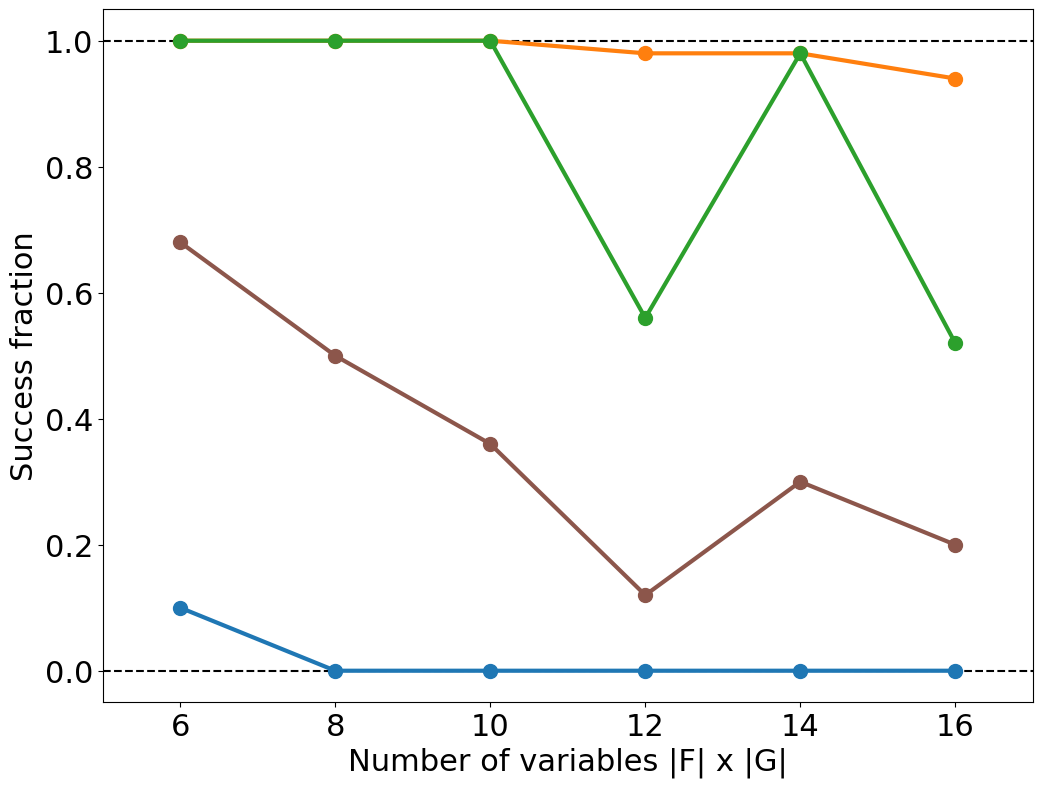}
        \caption{Fraction of successful runs for threshold $t=0.1$.}
        \label{fig:threshold_10}
    \end{minipage}
    \hfill
    \begin{minipage}{0.45\linewidth}
        \centering
        \includegraphics[width=0.97\linewidth]{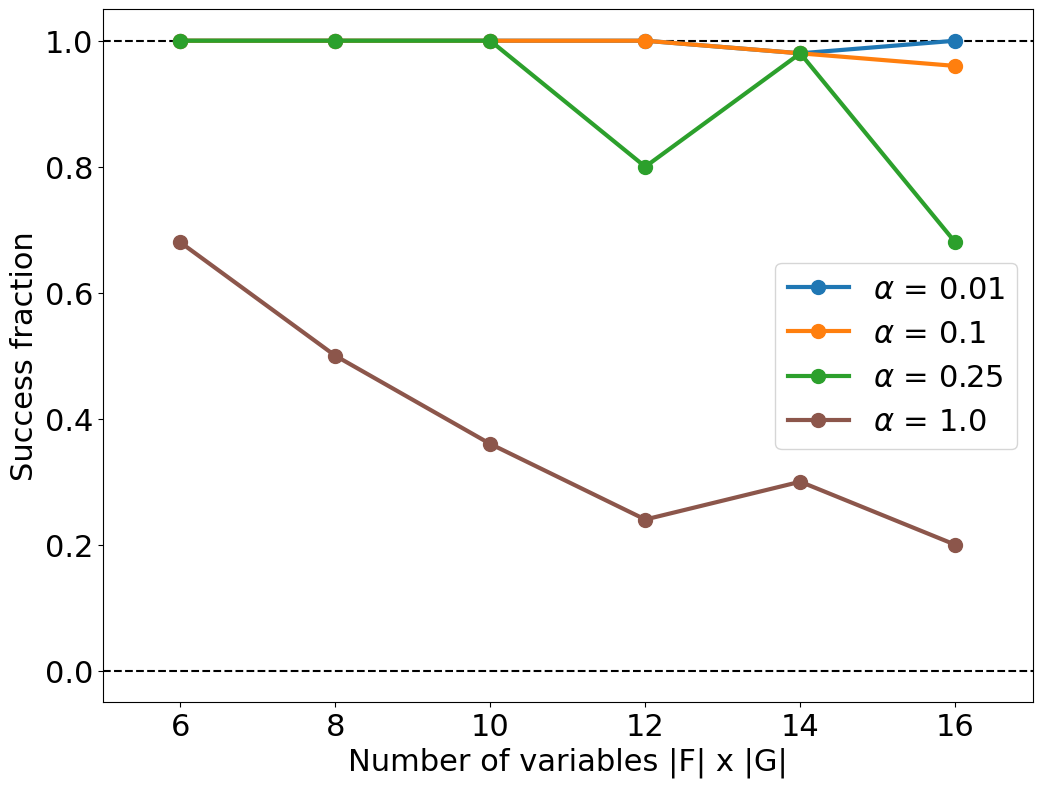}
        \caption{Fraction of successful runs for threshold $t=0.01$.}
        \label{fig:threshold_1}
    \end{minipage}
\end{figure*}

\section{The Flight-Gate Assignment problem}

The FGA problem aims at optimizing the assignment of flight activities $F$ to airport gates $G$. In this work, we consider a scenario where the total transfer time of passengers has to be minimized, subject to two linear constraints 1) each flight can be assigned to at most one gate 2) two flight activities cannot be assigned to the same gate simultaneously.

This problem can be formulated as a Quadratic Unconstrained Binary Optimization (QUBO) problem~\cite{Stollenwerk2019}
\begin{multline*}
    Q(\mathbf{x}) = T(\mathbf{x}) + \lambda^{\mathrm{one}}\sum_i \left( \sum_\alpha x_{i\alpha} - 1\right)^2 \\
    + \lambda^{\mathrm{not}}\sum_{(i,j) \in P} \sum_\alpha x_{i\alpha}x_{j\alpha} \, .
\end{multline*}
Here $\mathbf{x} = (x_{i\alpha})_{i\in F, \alpha\in G} \in\{0,1\}^{|F|\times|G|}$ is a set of binary variables representing an assignment ($x_{i\alpha} = 1$ if and only if flight $i\in F$ is assigned to gate $\alpha\in G$), $T(\mathbf{x})$ is a quadratic functional accounting for the total passenger travel time and the last two terms encode the two constraints, with $P\subset F\times F$ listing the incompatible pairs of flights and $\lambda^{\mathrm{one}}, \lambda^{\mathrm{not}} > 0$ are constants chosen large enough.

We can embed this problem into GBS with $N=|F|\times |G|$ optical modes as follows: the QUBO is turned into a Hamiltonian $\hat{Q}$ by replacing each binary decision variable $x_{i\alpha}$ with the projection $\hat{\Pi}^{(1)}_{i\alpha}$. The ground state of $\hat{Q}$ then encodes the optimal solution corresponding to the QUBO minimizer.

\section{VQE using the Conditional Value-at-Risk}

The \emph{Conditional Value-at-Risk} (CVaR) has been proposed to improve the VQE approach to combinatorial optimization problems by restricting the search space to the $\mathbf{x}$'s with the lowest energy~\cite{Barkoutsos2020}. Here, the cost function to minimize reads
\begin{align}
    \label{eq:cost_function}
    \mathcal{C}(\boldsymbol{\theta}) = \mathrm{CVaR}_\alpha(X(\hat{Q},\boldsymbol{\theta})), \quad \alpha\in (0,1] \, ,
\end{align}
where $X(\hat{Q},\boldsymbol{\theta})$ is the distribution of the observable $\hat{Q}$ in the quantum state $\ket{\varphi(\boldsymbol{\theta})}$ and $\mathrm{CVaR}_\alpha(X) = \mathbb{E}[X~|~X \leq F_X^{-1}(\alpha)]$ the CVaR with tail $\alpha$-left of a random variable $X$, and $F_X$ is the cumulative density function of $X$.

In the experiments, the cost function can be estimated by performing $K$ measurements of $\ket{\varphi(\boldsymbol{\theta})}$ and averaging over the $\lceil\alpha K \rceil$ lowest energy values. When $\alpha=1$, we recover the VQE cost function.

\section{Method}

We assess the performance of the GBS ansatz for solving the QUBO-FGA problem using the CVaR-VQE approach. The GBS is parameterized by the symmetric matrix $\boldsymbol{\theta} = U \mathrm{diag}(r_1,\dots, r_N) U^T$.
For fastening the training procedure, we train only the real parts of the $3N$ parameters corresponding to the smallest entries of the QUBO Hamiltonian.

The classical minimization is performed with constrained optimization by linear approximation (COBYLA), with at most $50 N$ function evaluations. We consider the randomly generated non-trivial and classically-hard problem instances in~\cite{Chai2023}. For each problem sizes $N \in \{6, 8, 10, 12, 14, 16\}$, we consider 50 different instances and run 5 training with random initialization. When $\alpha = 1$, the cost function in \eqref{eq:cost_function} admits an analytical expression with respect to $\boldsymbol{\theta}$. Therefore, we can use the highly efficient ADAM optimizer routines available in TensorFlow to minimize it.

We say an instance is \emph{successful} if one of the five runs results in a fidelity of the quantum state $\ket{\varphi(\boldsymbol{\theta}_*)}$ after training with the ground state of $\hat{Q}$ higher than a threshold $t \in \{0.1, 0.01\}$.

\section{Numerical results and comments}

The fraction of successful runs for fidelity threshold of $0.1$ and $0.01$ are presented respectively in Fig.~\ref{fig:threshold_10} and Fig.~\ref{fig:threshold_1}.

We observe that employing CVaR outperforms VQE significantly, achieving almost perfect success rates for $\alpha=0.01$ or $0.1$. The only exception is when $(\alpha, t) = (0.01, 0.1)$, which is attributed to the CVaR cost function's inability to reward fidelity higher than $\alpha$~\cite{Barkoutsos2020}. For $\alpha=0.25$ and $N=\{12,16\}$, the fraction of successful runs decreases. In these cases, the failing instances typically involve 3 or 4 flight activities. The algorithm appears here to be trapped in local minima.

\section{Conclusion and outlook}

In this study, we have tackled the FGA problem by employing GBS in combination with a CVaR-VQE approach, which significantly enhances performance compared to VQE. Despite focusing on small-sized instances, this work serves as an initial proof of principle for this new approach.

However, we believe that there is substantial room for improvement in the algorithm, for instance through careful choice of parameters to optimize. Another direction would use the binary encoding of the FGA problem, as proposed in~\cite{Chai2023}, which directly represents the first constraint.

Lastly, it would be beneficial to compare the use of the GBS ansatz with other ansatz types, specifically the one proposed in~\cite{Chai2023}. This comparison could offer valuable insights and further enhance our understanding of the problem-solving capabilities of different ansatz strategies.

\end{document}